\documentclass[fleqn,usenatbib]{mnras}

\usepackage{newtxtext,newtxmath}

\usepackage[T1]{fontenc}

\DeclareRobustCommand{\VAN}[3]{#2}
\let\VANthebibliography\thebibliography
\def\thebibliography{\DeclareRobustCommand{\VAN}[3]{##3}\VANthebibliography}

\usepackage{graphicx}	
\usepackage{amsmath}	

\newcommand\cm{{\rm\thinspace cm}}

\newcommand\Msun{\hbox{$\rm\thinspace M_{\odot}$}}

\newcommand\pc{{\rm\thinspace pc}}

\newcommand\yr{{\rm\thinspace yr}}

\newcommand\Hz{{\rm\thinspace Hz}}

\newcommand\pcmcu{\hbox{$\cm^{-3}\,$}}

\newcommand\Msunpyr{\hbox{$\Msun\yr^{-1}\,$}}

\title[Accretion discs at high-z ]{The possible accretion discs of GN-z11 at redshift $z=10.6$, MoM-z14 at $z=14.44$ and other high redshift objects}

\author[A. C. Fabian et al.]{
A. C. Fabian,$^{1}$\thanks{E-mail: acf@ast.cam.ac.uk },  J. Jiang$^2$,  W.M. Baker$^3$, R. Maiolino$^{4,5,6}$, X. Ji$^{4,5}$, I. Juodžbalis$^{4,5}$, J. Scholtz$^{4,5}$
\\
$^{1}$Institute of Astronomy, University of Cambridge, Madingley Road, Cambridge CB3 0HA, UK\\
$^2$Department of Physics, University of Warwick, Gibbet Hill Road, Coventry CV4 7AL, UK\\
$^3$DARK, Niels Bohr Institute, University of Copenhagen, Jagtvej 155A, DK-2200 Copenhagen, Denmark\\
$^{4}$Kavli Institute for Cosmology, University of Cambridge, Madingley Road, Cambridge, CB3 0HA, UK\\
$^{5}$Cavendish Laboratory, University of Cambridge, 19 JJ Thomson Avenue, Cambridge, CB3 0HE, UK\\
$^{6}$Department of Physics and Astronomy, University College London, Gower Street, London WC1E 6BT, UK\\
}
\date{Accepted XXX. Received YYY; in original form ZZZ}

\pubyear{2024}

\begin{document}
\label{firstpage}
\pagerange{\pageref{firstpage}--\pageref{lastpage}}
\maketitle

\begin{abstract}
The JWST has enabled the discovery of  Active Galactic Nuclei at high redshifts. The intrinsic UV spectrum of GN-z11 at redshift $z=10.6$ has a spectral slope compatible with a standard accretion disc. By fitting a disc model to its spectrum, we find that the mass of the black hole must be above $1.12\times 10^7\Msun$ in order that it lies below the Eddington limit. We define this mass as the Eddington mass of the black hole. We note that the spectral shape is consistent with that of accreting stellar mass black holes sources in their soft state, for which no variability is expected. Mom-z14 is a more distant object at $z=14.44$ and  has a similar UV slope. Disc model-fitting gives a similar result but lower mass accretion rate. We also examine
3 further high redshift objects: GS z14-1, GHZ2 and PAN-z14-1  at $z=13.86$, $12.34$  and $13.53$,  again obtaining similar results. 
If sub-Eddington accretion discs are indeed the origin of much of the UV emission from these objects, then the existence of massive black holes less than 304 and 290 Myr after the Big Bang point either to exceptional black hole seeds or to primordial black holes. The observed spread of UV spectral slopes in high redshift objects suggests that our approach may be relevant to about half of that population.

\end{abstract}




\section{Introduction}

The James Webb Space Telescope (JWST) has opened a totally new discovery space in the exploration of black hole accretion in the early Universe. Indeed, the unprecedented sensitivity of its data, together with the extended wavelength range, have revealed a large population of Active Galactic Nuclei (AGN) much fainter than quasars found from previous groundbased observations, as well as AGN at much higher redshift \citep[e.g.][]{Harikane2023,Maiolino2024JADES,Kocevski2025,Taylor2025,Juodzbalis2025}. Among these,
GN-z11, at redshift z = 10.6, is the most distant AGN found so far \citep{maiolino24}.
It was first identified with HST \citep{Oesch2016} and then observed extensively with JWST. 
\cite{Bunker23}
 used a first spectrum of
GN-z11 in the context of a powerful star-forming galaxy. 
\cite{Tacchella2023}
showed that the morphology consists of a compact (essentially unresolved)
source, surrounded by a marginally extended component, accounting for about
30\% of the UV light.
Subsequently, \cite{maiolino24} used deeper JWST spectroscopy to show that
some of the UV permitted lines  (NIV]1486 and NIII]1750) point to gas densities
in excess of $10^9\pcmcu$, typical of the Broad Line Region (BLR) of AGN.
They argue that the presence of an AGN dominating the central compact source is
also supported by the presence of the CII$^*$1335 line (typical of quasars),
the high ionization line [NeIV]2424, and a fast outflow identified in CIV1549. They
already noticed that the UV spectral slope is consistent with that of standard accretion
discs ($\beta \approx -2.33$) and the nebular emission of GN-z11 is consistent with negligible dust attenuation. \cite{Scholtz2024GNz11} used JWST-NIRSpec IFU data to reveal a luminous and extended
Ly$\alpha$ halo consistent with that observed in quasars (and inconsistent with the fainter
Ly$\alpha$ haloes found in star forming galaxies), and \cite{Maiolino2024PopIII} also showed
that the extended CIII] morphology is fan-shaped, as seen in AGN. \cite{Ji2025GNz11}
also
found the signature of a UV bump, which they interpret as FeII emission, similar to that seen in type 1 AGN.

On the other hand, \cite{Alvarz-Marquez2025} presented MIRI/MRS spectroscopy of GN-z11, revealing narrow [OIII]5007
and H$\alpha$ (although an underlying broad H$\alpha$ is not ruled out), and favouring excitation by star formation.
Also, \cite{Gunawardhana_gnz11_2025} argue the observed line ratios of GN-z11 could be explained by ionization of stellar populations, including Wolf-Rayet stars.
However, one should take into account
that the new population of AGN discovered by JWST, even if unambiguously identified as type 1, is undistinguishable from metal-poor and high-ionisation star forming galaxies
in terms of optical nebular diagnostics, as extensively demonstrated by multiple studies 
\citep[e.g.][]{Uebler2023,Maiolino2024JADES,Juodzbalis2025}.
More recently, Maiolino et al. (in prep.) have used an ultradeep co-added spectrum of GN-z11 to further
confirm the AGN nature of its case, by accurately measuring the nuclear gas density as
$10^{10}\pcmcu$, unambiguously proving the presence of a BLR. Contribution by Wolf-Rayet stars has been proposed in the past
\citep{Senchyna2024}, however this
has been later excluded based on the absence of NIV1718 
\citep[which should be very strong in WN stars,][]{maiolino24}
and the absence of a broad component of HeII1640 (Maiolino et al. in prep.), which is the key WR
signature.

\begin{figure*}
    \centering
    \includegraphics[width=\textwidth]{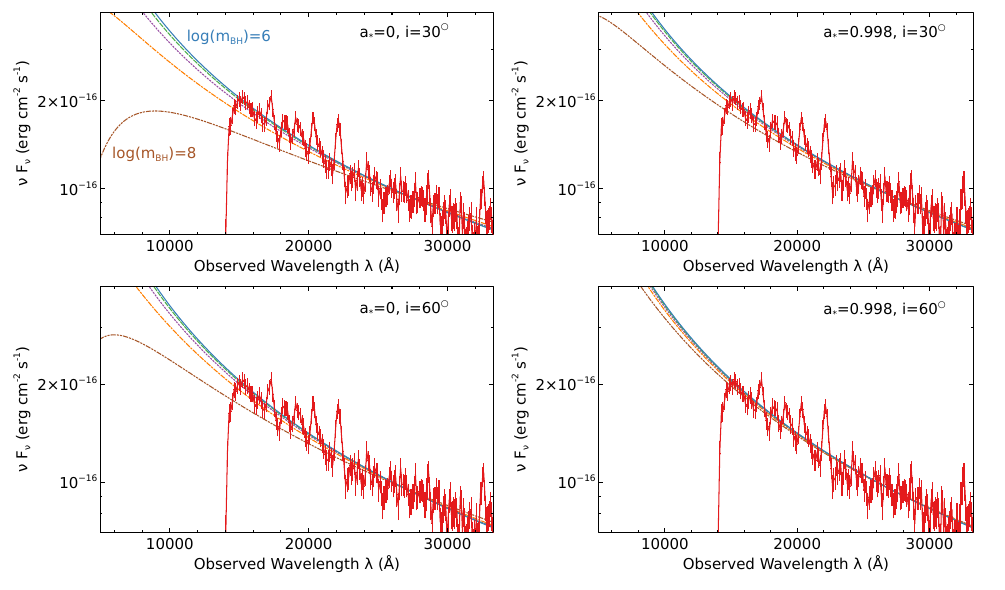}
    \caption{ The NIRSpec spectrum of GN-z11 fitted with a relativistic multi-colour disc thermal emission model assuming different combinations of BH spin and inclination angle. The solid, dashed, dotted, dash-dotted and dash-dot-dotted lines assume $\log(m_{\rm BH}) = $ 6, 6.5, 7, 7.5 and 8 respectively. The $\log(m_{\rm BH}) = $7.5 model deviates from the observed spectrum below 20,000 \AA\ in the top left panel for a low-spin and low-inclination angle model. In the bottom right panel, for the maximum BH spin and a high inclination angle, the difference between these models for different BH masses is very small in the observed wavelength range. A high-spin and high-inclination angle model would allow a wider range of possible BH masses to fit the data.}
    \label{pic_jwst_fit3}
\end{figure*}

\section{Spectral Results for GN-z11}
\begin{figure}
    \centering
    \includegraphics[width=\linewidth]{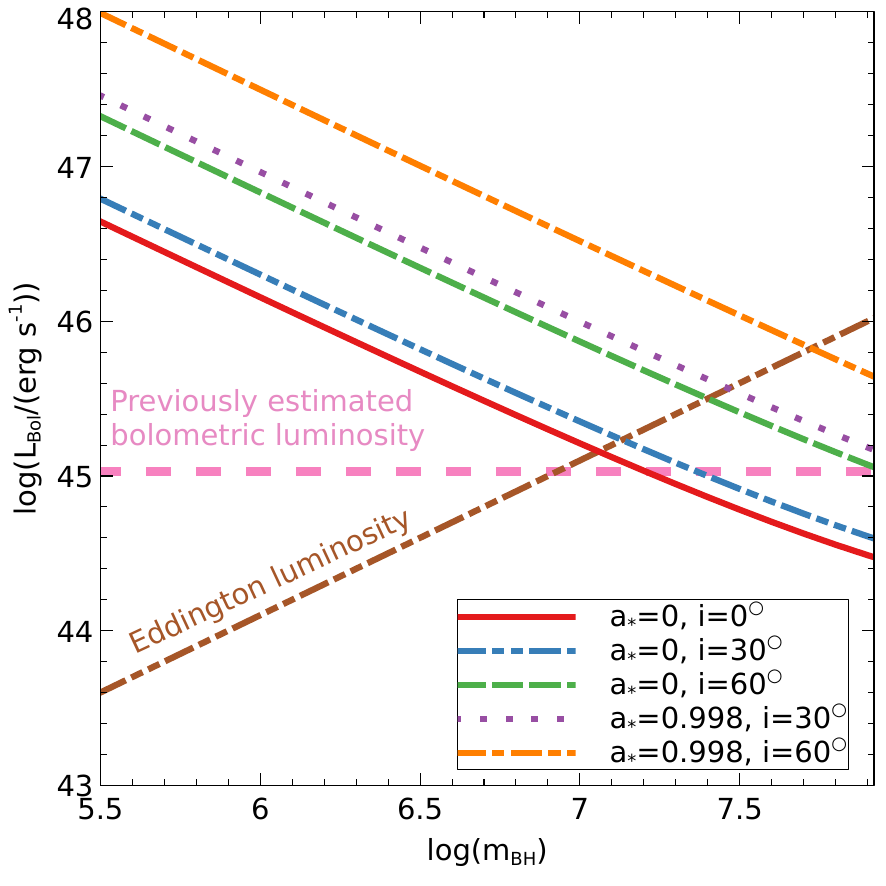}
    \caption{The inferred bolometric luminosity of the multi-colour disc thermal emission assuming different combinations of BH spin and disc inclination angle in comparison with Eddington luminosity (dash-dot-dotted line). The pink horizontal dashed line shows the bolometric luminosity of GN-z11 calculated by applying a correction factor to the rest-frame 1400~\AA\ luminosity of the object \citep{maiolino24}. \\(Note: $L_{\rm Bol}=\epsilon\dot{M}c^{2}$ in this calculation, with $\epsilon=0.057$. The Eddington BH mass is $\log(m_{\rm BH})=7.05$ (or $1.12\times10^{7}M_{\odot}$) for $a=0$ and $i=30$.)}
    \label{pic_lum2}
\end{figure}

Here, we model the residual 70\% continuum of GN-z11 as a blackbody accretion disc \citep{Lynden-Bell1969,SS73, NT73} and use it to deduce limits on the mass and spin of the central black hole. We also compare its luminosity to the Eddington limit. Such an accretion disc has the same shape irrespective of black hole mass, which only causes a shift in its flux and the turnover frequency due to the innermost edge (i.e. the Innermost Stable Circular Orbit or ISCO), despite a millionfold or more difference in black hole mass from stellar mass system to AGN. The power-law part of the spectrum has an index of $-7/3$ in a plot of flux $F_{\lambda}$ versus wavelength $\lambda$ and it is $-1/3$ if versus frequency $\nu$. If $F_{\nu}$ against $\nu$ is used then it scales as $F_\nu\propto{\nu}^{1/3}$.
The observed slope reported by \cite{maiolino24} is therefore compatible with that from a luminous accretion disc around stellar mass black holes in the soft state. Interestingly, if this is the correct identification, then there should be no  rapid variability or X-ray corona, as indeed observed so far for GN-z11. 

Deducing black hole masses from the spectra of their accretion discs started with 3C273 in 1978 \citep{Shields78} and progressed  with the work of \cite{Kriss99}, \cite{Davis11}, \cite{Raimundo12}, \cite{Calderone13}, and \cite{Campitiello18}, to cite a few. The later ones used approximations based on the \textsc{Kerrbb} model \citep{Li05} of the Novikov-Thorne disc \citep{NT73}. Here we fit the observed spectrum of GN-z11 directly with the \textsc{Kerrbb} model.

We use XSPEC v12.14.1 \citep{arnaud96} for the spectral fitting.  We exclude the wavelength ranges where emission lines are reported in \citet{maiolino24} and the flux is reduced to 70 per cent to account for the host galaxy and nebular emission (as for the spectrum shown in the right panel of Fig 9  in that paper). 
The disc is modelled with the relativistic thermal emission model \texttt{zkerrbb}\footnote{For reproducibility, note that versions of \texttt{zkerrbb} before XSPEC 12.13 underestimated the flux by a factor of $(1+$z$)^{2}$.} \citep{Li05}.  
The disc emission is limb-darkened and includes returning radiation.  
We use a spectral hardening factor of 1.7.  
We test different black hole masses ($m_{\rm BH}=M_{\rm BH}/M_{\odot}$), disc inclinations $i$, and spins $a_{*}$.  
Table\,\ref{tab_fit} lists examples of best-fit models.  
Their spectra are shown in Fig.\,\ref{pic_jwst_fit3}.  
Most fits have $\chi^{2}\approx230$ with 225 degrees of freedom (DOF).  
For $a_{*}=0$ and $i=30^{\circ}$, a disc with $m_{\rm BH}=10^{8}$ fits worse, with $\chi^{2}=432.22$.  
This is because the model underestimates flux below 2000\,\AA\ due to a lower disc temperature (dash-dot-dotted line in the top left panel of Fig.\,\ref{pic_jwst_fit3}).

\begin{table}
    \centering
    \begin{tabular}{ccccc}
    \hline\hline
    $a_{*}$ & $i$ & $\log(m_{\rm BH})$ & $\dot{m}$ & $\chi^{2}/$DOF \\
    - & (deg) & - & ($M_{\odot}$/yr) & - \\
    \hline
    0 & 30 & 7 & $0.696\pm0.010$ & 230.27/225\\
    0 & 30 & 6 & $6.20\pm0.10$ & 233.71/225 \\
    0 & 30 & 8 & $0.108\pm0.001$ & 432.22/225 \\
    0 & 60 & 7 & $2.29\pm0.04$ & 229.87/225 \\
    0.998 & 30 & 7 & $0.547\pm0.009$ & 230.29/225 \\
    \hline\hline
    \end{tabular}
    \caption{Best-fit mass accretion rates for GN-z11 are found for different $a_{*}$, $i$, and $m_{\rm BH}$.  
A high mass $m_{\rm BH}>10^{8}$, with low spin $a_{*}=0$ and low inclination $i=30^{\circ}$, gives a worse fit 
since the model spectrum turns over at a lower temperature,  reducing the predicted flux below observed 20,000 \AA. 
}
    \label{tab_fit}
\end{table}

It is well-known that the simple accretion disc model does not fit the UV spectra of quasars well \citep{Davis07}. The likely explanation is intervening, mild dust absorption \citep{Davis07, Temple21}. The spectral agreement found here in GN-z11 is probably due to GN-z11 having little dust \citep{Bunker23,Alvarz-Marquez2025}. This issue is briefly explored further in the Appendix.

In order to be sub-Eddington the mass has to be at least   $1.12\times10^{7}M_{\odot}$ for $a_{*}=0$ and $i=30^{\circ}$ (Fig. 2). This disagrees with the mass reported from the JWST spectra \citep{maiolino24}, which used the width of the high S/N and isolated NIV line  to obtain a black hole mass of $1.6\times 10^6\Msun$ with an uncertainty of a factor of about 2, which would result in a conclusion of super-Eddington accretion. The bolometric luminosity was then estimated using the 1400\AA\ bolometric corrections presented by \citep{Netzer19}, compiled using a relativistic disc model as used in the present work, and scaled for the above low mass\footnote{The \citep{Netzer19} bolometric corrections are averaged described as an "eye-fitted approximation" over a luminosity range and as such are mass and spin independent}. Our more massive solutions can easily have a sub-Eddington luminosity.

\section{A possible accretion disc in MoM-z14 and two other high redshift objects?}

A luminous object at $z=14.44$ has recently been discovered with JWST by  \cite{Naidu25}. It is extremely compact with an effective radius of $74\pc$. Any extension may be from the host galaxy, and an AGN can still be a significant source of UV light. It does have a steep UV slope of $-2.5\pm0.2$ which is consistent with that of an accretion disc ($-2.33$) as for GN-z11.  Here we consider the possibility that 70 per cent originates in an accretion disc and rescale it to the GN-z11 results above. We adopt the host-subtracted photometry and uncertainties from Table 2 of \citep{Naidu25}, both scaled by a factor of 0.7 to account for the AGN contribution.

Rather than fit the JWST spectrum of MoM-z14 directly, we scale its rest-frame luminosity to the models obtained above  for GN-z11. The luminosity scales as
\begin{equation}
  L_{\nu} \propto  {\nu}^{1/3} (m \dot m)^{2/3} 
\end{equation}
so
\begin{equation}
  \dot m \propto (\nu L_{\nu} )^{3/2} \nu^{-2} m^{-1}
\end{equation}
and
\begin{equation}L_{\rm bol}\propto \eta \dot m c^2 
\end{equation}
which equals the Eddington Limit $L_{\rm Edd}$ when
\begin{equation}
 m\propto\sqrt \eta (\nu L_{\nu} )^{3/4} \nu^{-1}.   
\end{equation}
The GN-z11 result for $a=0$, $i=30$, $\eta=0.057$, is

\begin{equation}
m_{\rm Edd}= 1.12\times10^7
\left(\frac{\nu L_{\nu}}{10^{44.3}}\right)^{3/4}
\left(\frac{\nu}{10^{15.33}}\right)^{-1}
\,M_\odot
\end{equation}

This generalises to 

\begin{equation}
m_{\rm Edd}= 4.4\times10^6
\left(\frac{\nu L_{\nu}}{10^{44}}\right)^{3/4}
\left(\frac{\nu}{10^{15.33}}\right)^{-1}
\,M_\odot
\end{equation}

This gives  $m_{\rm Edd}$ for Mom-z14 of $8.2\times 10^6\Msun$ using 44.12 for $\nu L_{\nu}$ from \citep{Naidu25}. We use a wavelength of 1400\AA\ ($\nu=2.14\times 10^{15}\Hz$). 

We define $m_{\rm Edd}$ as the Eddington mass for an object. For a given UV luminosity from the -7/3 power-law part of an accretion disc spectrum, it is the minimum mass that does not exceed the Eddington limit. We use $i=30^{\circ}$ since exactly face-on is unlikely.

If the object has a prograde spin then its bolometric luminosity increases due to the shrinkage of the ISCO. The mass required to be at the Eddington limit then also increases, as shown in Fig.~\ref{pic_lum2}.

\begin{figure*}
    \centering
    \includegraphics[width=\textwidth]{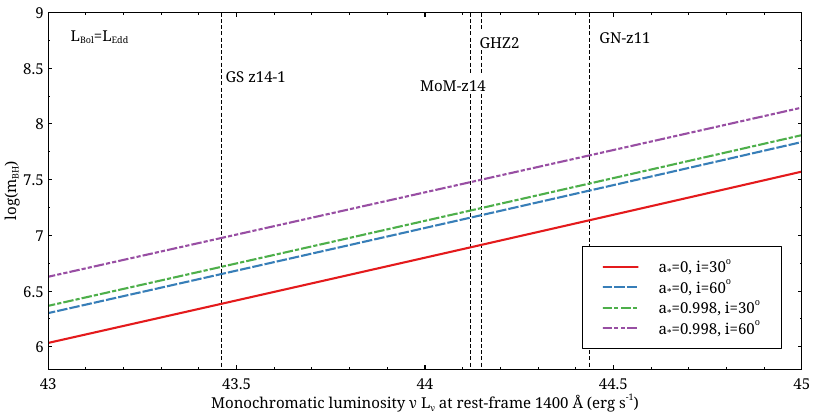}
    \caption{Monochromatic luminosity at rest-frame 1400\,\AA\ in erg\,s$^{-1}$ for different black hole masses $m_{\rm BH}$ inferred by the \texttt{zkerrbb} model. Lines with different styles show different parameter sets. The calculations assume the bolometric luminosity of the accretion disc equals the Eddington limit. The vertical dashed line marks the monochromatic luminosity of several JWST-observed, high-redshift objects at rest-frame 1400\,\AA. This figure can be used to estimate black hole mass from a photometric luminosity, without prior knowledge of the spectral shape.}
    \label{pic_mbh_pre2}
\end{figure*}

\begin{table*}
    \centering
    \begin{tabular}{ccccccccc}
    \hline\hline
      Sources & $z$ & $\log(\nu L_{\nu})$ & \textcolor{red}{UV Slope} & Ref & $\log(m_{\rm BH})$ & $\dot{m}$& $t=m/\dot m$ & Age\\
      - & - & erg s$^{-1}$ & $\beta$ & - & - & $M_{\odot}$ yr$^{-1}$ & Myr & Myr\\
      \hline
      GN-z11 & 10.6 & 44.3 & $-2.26 \pm 0.1$ & \citet{maiolino24} & 7.05  & 0.44 & 26 & 441 \\
      MoM-z14  & 14.44 & 44.12 & $-2.5\pm0.2$ &\citet{Naidu25} & 6.9 & 0.31 & 26 & 290\\
      GS z14-1 & 13.86 & 43.46 & $-2.32\pm0.08$ & \citet{Wu2025} & 6.4 & 0.10 & 25 & 304\\
      GHZ2 & 12.34 & 44.15 & $-2.46\pm0.08$& \citet{castellano24} & 6.9 & 0.29 & 27 & 380\\
      \hline\hline
    \end{tabular}
    \caption{Derived black hole masses for several high-$z$ JWST-observed objects.  
The third column lists their monochromatic luminosity at rest-frame 1400\,\AA, obtained by visually measuring the flux density from the spectra given in the references in the fourth column.  
The sixth column gives Eddington masses inferred from a relativistic multi-colour disc model (\texttt{zkerrbb}), assuming a bolometric disc luminosity at the Eddington limit, spin $a_{*}=0$, and inclination $i=30^{\circ}$.  
Other choices, such as a lower bolometric luminosity, higher spin, or higher inclination, would lead to larger black hole mass estimates. The final 3 columns list the mass accretion rate, mass doubling time and the age of the Universe at the redshift of the object. 
}
    \label{tab:placeholder}
\end{table*}

Another distant compact, faint object with a similar UV spectrum, JADES-GS-z14-1, has  been reported at $z=13.86$ by \citep{Wu2025}. The authors discuss briefly an AGN origin and conclude that it might have a black hole mass of about $10^6\Msun$ or less if it is in a super-Eddington state. It is about 2.5 times fainter than MoM-z14 at the same UV wavelengths so its mass accretion rate is lower by a factor of about 4. We suspect that it may be sub-Eddington and then has $m\sim 10^7\Msun$ like Mom-z14. It also has a low metallicity.

A further  high redshift object, GHZ2,  \citep{ castellano24}, discovered by JWST shows an appropriate  UV slope. The Eddington mass for its black hole has been obtained from its published UV spectrum {see Table 2 and  Fig.3}. We note that the emission line diagnostics of  GHZ2 are consistent with both AGN and low metallicity star-formation, 

One new object at $z=13.53$, PAN-z14-1 \citep{Donnan26} has a UV slope of $\beta=-2.26\pm 0.08$, which matches an accretion disc very well. Its UV flux  is similar to that of Mom-z14 so, from our prescription, its mass will be similar ($\sim 10^7\Msun$). It does however appear to be extended with a circularized half-light radius $r_c=233\pm 10\pc$. We suggest that this could indicate multiple accreting  black holes in this object, all of which will have the same spectral shape but different luminosities. Note that observations of triple massive black hole systems have been reported by \citep{ubler25, Schwartzman25} and references therein.

\section{Discussion}

We have modelled the UV spectra of 4 of the most distant objects yet found by JWST with a standard accretion disc spectrum. This has enabled us to place constraints on their mass accretion rates. Assuming that these rates do not exceed the Eddington value we obtain Eddington masses of $10^{6.4}$ to $10^{7.05}\Msun$ for them. If they are spinning rapidly, then their true masses  and/or the inclination must be higher for them to be below  the Eddington limit (Fig. 2). Their low-spin mass accretion rates range from  0.7 and $0.1\Msunpyr$. 

As defined above, the Eddington mass $m_{\rm Edd}$ is the minimum mass consistent with a UV spectral index of $\beta=-2.33$. The actual mass of the black hole  may be less than $m_{\rm Edd}$ but it must then be accreting at a super-Eddington rate. The inner disc will then be puffed up, possibly irradiating the outer disc more; it may also have mass outflows and the radiation may be beamed. All these effects can plausibly make the UV slope depart from -2.33 \citep{Madau25}. Furthermore, there are also environmental issues, such as dust and possible nebular continuum to contend with. \cite{Cheng19} discuss departures from the standard accretion disc slope. Response to slowly declining or rising accretion rates could be relevant to young AGN. The UV slope criterion is important for ensuring that we are using sub-Eddington accretors when applying our formalism.  

\cite{Napolitano25} discuss the UV slope $\beta$ for 24 galaxies at $z>9$. About half of them have $-2<\beta>-2.5$ when uncertainties are considered, indicating that our approach may be applied to about half the population.  This is consistent with recent findings from \citet{Geris2025} which identified a population of low mass AGN in the stacks of JWST/NIRspec data.  {\citet{Tang25} study JWST spectroscopic properties of high redshift galaxies and note that the intrinsic slope set by stellar population and ionization conditions gives $\beta=-2.6$ to $-2.4$ before reddening. They have 14 sources in the redshift range $z=11-14$ giving an average $\beta$ of $-2.37^{+0.05}_{-0.09}$, which compares well with the accretion disc slope of $-2.33$.

Two high redshift galaxies, GHZ9 at $z=10.145$ \citep{Kovacs24, Alvarz-Marquez2025} and UHZ1 \citep{Bogdan24} $z=10.3073$ \citep{Goulding23, Alvarz-Marquez2025}, have reported hard X-ray emission indicating coronal emission from accretion and thus supermassive black holes. The first has a UV slope of $-1.10\pm0.12$ and the second cannot contain a luminous absorbed AGN \citep{CrespoGomez25}. Therefore  We do not  apply our technique to them. There is however mounting evidence for AGN from JWST spectroscopy in  other high redshift objects \citep{Zhu26, Chavez25, Castellano25}.

Thermal accretion discs in soft state X-ray binary systems that are  fit well  by \textsc{kerrbb} do not significantly exceed the Eddington limit\footnote{We ignore any jetted emission} and we have no reason to suppose that the situation is different for supermassive black holes.

The black hole mass-doubling timescales of 16 to 27 Myr are short compared  with the ages since the Big Bang of  441 to 290 Myr.    We have used our model fitting results to map out the AGN luminosity--Black Hole Mass plane (Fig 4) for GN-z11. 

These results point at least to heavy black hole seeds, or quasistars \citep{Coughlin24} in the case of GN-z11. In the low metallicity cases  of MoM-z14 and JADES-GS-z14-1, primordial black hole seeds such as proposed for the exceptionally low metallicity quasar Abell-2755-QSO1 at $z=7.04$ \citep{Maiolino25}, may be required. 
If primordial black hole seeds are responsible, then there must be a substantial population of  high redshift, $z>10$, AGN yet to be discovered \citep{Dayal25}. {Multiple black hole systems may also  be common in this scenario.}

\begin{figure}
    \centering
    \includegraphics[width=1.0\linewidth]{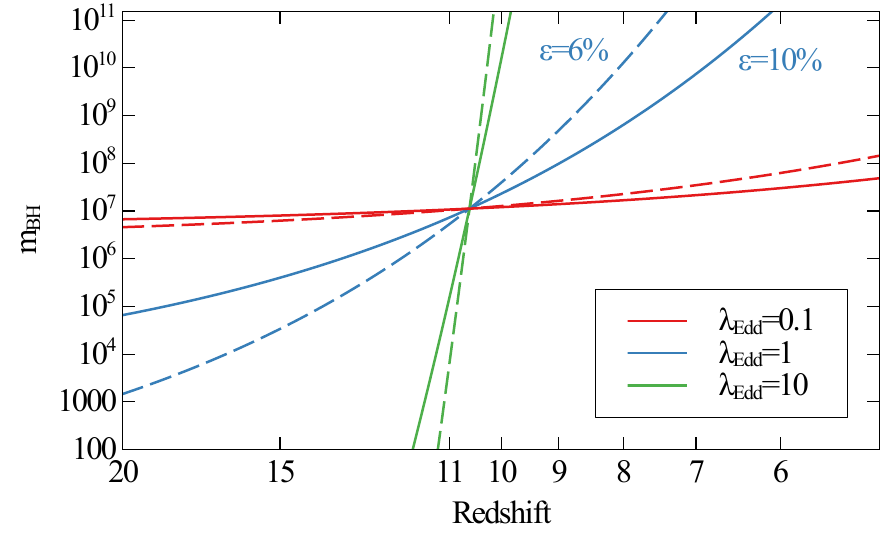}
    \caption{Black hole mass ($m_{\rm BH}$) plotted against redshift for 3 values of Eddington fraction ($\lambda_{\rm Edd}= 0.1, 1\ {\rm and}\ 10$ red, blue and green respectively and radiative efficiency of $6\%$ dashed and $10\%$ solid). The lines are anchored at $\log(m)=7.05$, the Eddington mass of GN-z11.}
    \label{pic_edd2}
\end{figure}

\section*{Acknowledgements}
We thank the referee for helpful comments and Sandro Tacchella for a discussion.
WMB gratefully acknowledges support from DARK via the DARK fellowship. This work was supported by a research grant (VIL54489) from VILLUM FONDEN.

\section*{Data Availability}
There are no new data associated with this article.
All JWST data used in this article can be accessed through MAST.



\bibliographystyle{mnras}
\bibliography{gnz11disc} 

@ARTICLE{Geris2025,
       author = {{Geris}, Sophia and {Maiolino}, Roberto and {Isobe}, Yuki and {Scholtz}, Jan and {D'Eugenio}, Francesco and {Ji}, Xihan and {Juodzbalis}, Ignas and {Simmonds}, Charlotte and {Dayal}, Pratika and {Trinca}, Alessandro and {Schneider}, Raffaella and {Arribas}, Santiago and {Bhatawdekar}, Rachana and {Bunker}, Andrew J. and {Carniani}, Stefano and {Charlot}, Stephane and {Chevallard}, Jacopo and {Curtis-Lake}, Emma and {Johnson}, Benjamin D. and {Parlanti}, Eleonora and {Rinaldi}, Pierluigi and {Robertson}, Brant and {Tacchella}, Sandro and {Uebler}, Hannah and {Venturi}, Giacomo and {Williams}, Christina C. and {Witstok}, Joris},
        title = "{JADES reveals a large population of low mass black holes at high redshift}",
      journal = {arXiv e-prints},
         year = 2025,
        month = jun,
          eid = {arXiv:2506.22147},
        pages = {arXiv:2506.22147},
          doi = {10.48550/arXiv.2506.22147},
archivePrefix = {arXiv},
       eprint = {2506.22147},
 primaryClass = {astro-ph.GA},
       adsurl = {https://ui.adsabs.harvard.edu/abs/2025arXiv250622147G}
}

@ARTICLE{Gunawardhana_gnz11_2025,
       author = {{Gunawardhana}, Madusha L.~P. and {Brinchmann}, Jarle and {Croom}, Scott and {Bunker}, Andrew and {Bryant}, Julia and {Oh}, Sree},
        title = "{JADES NIRSpec Spectroscopy of GN-z11: Evidence for Wolf-Rayet contribution to stellar populations at 430 Myr after Big Bang?}",
      journal = {arXiv e-prints},
         year = 2025,
        month = apr,
          eid = {arXiv:2504.12584},
        pages = {arXiv:2504.12584},
          doi = {10.48550/arXiv.2504.12584},
archivePrefix = {arXiv},
       eprint = {2504.12584},
 primaryClass = {astro-ph.GA},
       adsurl = {https://ui.adsabs.harvard.edu/abs/2025arXiv250412584G}
}

@ARTICLE{maiolino24,
       author = {{Maiolino}, Roberto and {Scholtz}, Jan and {Witstok}, Joris and {Carniani}, Stefano and {D'Eugenio}, Francesco and {de Graaff}, Anna and {{\"U}bler}, Hannah and {Tacchella}, Sandro and {Curtis-Lake}, Emma and {Arribas}, Santiago and {Bunker}, Andrew and {Charlot}, St{\'e}phane and {Chevallard}, Jacopo and {Curti}, Mirko and {Looser}, Tobias J. and {Maseda}, Michael V. and {Rawle}, Timothy D. and {Rodr{\'\i}guez del Pino}, Bruno and {Willott}, Chris J. and {Egami}, Eiichi and {Eisenstein}, Daniel J. and {Hainline}, Kevin N. and {Robertson}, Brant and {Williams}, Christina C. and {Willmer}, Christopher N.~A. and {Baker}, William M. and {Boyett}, Kristan and {DeCoursey}, Christa and {Fabian}, Andrew C. and {Helton}, Jakob M. and {Ji}, Zhiyuan and {Jones}, Gareth C. and {Kumari}, Nimisha and {Laporte}, Nicolas and {Nelson}, Erica J. and {Perna}, Michele and {Sandles}, Lester and {Shivaei}, Irene and {Sun}, Fengwu},
        title = "{A small and vigorous black hole in the early Universe}",
      journal = {\nat},
         year = 2024,
        month = mar,
       volume = {627},
       number = {8002},
        pages = {59-63},
          doi = {10.1038/s41586-024-07052-5},
archivePrefix = {arXiv},
       eprint = {2305.12492},
 primaryClass = {astro-ph.GA},
       adsurl = {https://ui.adsabs.harvard.edu/abs/2024Natur.627...59M}
}

@ARTICLE{Harikane2023,
       author = {{Harikane}, Yuichi and {Zhang}, Yechi and {Nakajima}, Kimihiko and {Ouchi}, Masami and {Isobe}, Yuki and {Ono}, Yoshiaki and {Hatano}, Shun and {Xu}, Yi and {Umeda}, Hiroya},
        title = "{A JWST/NIRSpec First Census of Broad-line AGNs at z = 4-7: Detection of 10 Faint AGNs with M $_{BH}$ {}10$^{6}$-{}10$^{8}$ M $_{{\ensuremath{\odot}}}$ and Their Host Galaxy Properties}",
      journal = {\apj},
         year = 2023,
        month = dec,
       volume = {959},
       number = {1},
          eid = {39},
        pages = {39},
          doi = {10.3847/1538-4357/ad029e},
archivePrefix = {arXiv},
       eprint = {2303.11946},
 primaryClass = {astro-ph.GA},
       adsurl = {https://ui.adsabs.harvard.edu/abs/2023ApJ...959...39H}
}

@ARTICLE{Lynden-Bell1969,
       author = {{Lynden-Bell}, D.},
        title = "{Galactic Nuclei as Collapsed Old Quasars}",
      journal = {\nat},
         year = 1969,
        month = aug,
       volume = {223},
       number = {5207},
        pages = {690-694},
          doi = {10.1038/223690a0},
       adsurl = {https://ui.adsabs.harvard.edu/abs/1969Natur.223..690L}
}

@ARTICLE{Senchyna2024,
       author = {{Senchyna}, Peter and {Plat}, Adele and {Stark}, Daniel P. and {Rudie}, Gwen C. and {Berg}, Danielle and {Charlot}, St{\'e}phane and {James}, Bethan L. and {Mingozzi}, Matilde},
        title = "{GN-z11 in Context: Possible Signatures of Globular Cluster Precursors at Redshift 10}",
      journal = {\apj},
         year = 2024,
        month = may,
       volume = {966},
       number = {1},
          eid = {92},
        pages = {92},
          doi = {10.3847/1538-4357/ad235e},
archivePrefix = {arXiv},
       eprint = {2303.04179},
 primaryClass = {astro-ph.GA},
       adsurl = {https://ui.adsabs.harvard.edu/abs/2024ApJ...966...92S}
}

@ARTICLE{Uebler2023,
       author = {{{\"U}bler}, Hannah and {Maiolino}, Roberto and {Curtis-Lake}, Emma and {P{\'e}rez-Gonz{\'a}lez}, Pablo G. and {Curti}, Mirko and {Perna}, Michele and {Arribas}, Santiago and {Charlot}, St{\'e}phane and {Marshall}, Madeline A. and {D'Eugenio}, Francesco and {Scholtz}, Jan and {Bunker}, Andrew and {Carniani}, Stefano and {Ferruit}, Pierre and {Jakobsen}, Peter and {Rix}, Hans-Walter and {Rodr{\'\i}guez Del Pino}, Bruno and {Willott}, Chris J. and {Boeker}, Torsten and {Cresci}, Giovanni and {Jones}, Gareth C. and {Kumari}, Nimisha and {Rawle}, Tim},
        title = "{GA-NIFS: A massive black hole in a low-metallicity AGN at z {\ensuremath{\sim}} 5.55 revealed by JWST/NIRSpec IFS}",
      journal = {\aap},
         year = 2023,
        month = sep,
       volume = {677},
          eid = {A145},
        pages = {A145},
          doi = {10.1051/0004-6361/202346137},
archivePrefix = {arXiv},
       eprint = {2302.06647},
 primaryClass = {astro-ph.GA},
       adsurl = {https://ui.adsabs.harvard.edu/abs/2023A&A...677A.145U}
}

@ARTICLE{Alvarz-Marquez2025,
       author = {{{\'A}lvarez-M{\'a}rquez}, J. and {Crespo G{\'o}mez}, A. and {Colina}, L. and {Langeroodi}, D. and {Marques-Chaves}, R. and {Prieto-Jim{\'e}nez}, C. and {Bik}, A. and {Alonso-Herrero}, A. and {Boogaard}, L. and {Costantin}, L. and {Garc{\'\i}a-Mar{\'\i}n}, M. and {Gillman}, S. and {Hjorth}, J. and {Iani}, E. and {Jermann}, I. and {Labiano}, A. and {Melinder}, J. and {Meyer}, R. and {{\"O}stlin}, G. and {P{\'e}rez-Gonz{\'a}lez}, P.~G. and {Rinaldi}, P. and {Walter}, F. and {van der Werf}, P. and {Wright}, G.},
        title = "{Insight into the starburst nature of Galaxy GN-z11 with JWST MIRI spectroscopy}",
      journal = {\aap},
         year = 2025,
        month = mar,
       volume = {695},
          eid = {A250},
        pages = {A250},
          doi = {10.1051/0004-6361/202451731},
archivePrefix = {arXiv},
       eprint = {2412.12826},
 primaryClass = {astro-ph.GA},
       adsurl = {https://ui.adsabs.harvard.edu/abs/2025A&A...695A.250A}
}

@ARTICLE{Ji2025GNz11,
       author = {{Ji}, Xihan and {Maiolino}, Roberto and {Ferland}, Gary and {D'Eugenio}, Francesco and {Bhatawdekar}, Rachana and {Charlot}, St{\'e}phane and {Chevallard}, Jacopo and {Curti}, Mirko and {Curtis-Lake}, Emma and {Hainline}, Kevin and {Ji}, Zhiyuan and {Robertson}, Brant and {Rodr{\'\i}guez Del Pino}, Bruno and {Scholtz}, Jan and {Tacchella}, Sandro and {Williams}, Christina C. and {Witstok}, Joris},
        title = "{JADES {\textendash} the small blue bump in GN-z11: insights into the nuclear region of a galaxy at z = 10.6}",
      journal = {\mnras},
         year = 2025,
        month = aug,
       volume = {541},
       number = {3},
        pages = {2134-2161},
          doi = {10.1093/mnras/staf1083},
archivePrefix = {arXiv},
       eprint = {2405.05772},
 primaryClass = {astro-ph.GA},
       adsurl = {https://ui.adsabs.harvard.edu/abs/2025MNRAS.541.2134J}
}

@ARTICLE{Maiolino2024PopIII,
       author = {{Maiolino}, Roberto and {{\"U}bler}, Hannah and {Perna}, Michele and {Scholtz}, Jan and {D'Eugenio}, Francesco and {Witten}, Callum and {Laporte}, Nicolas and {Witstok}, Joris and {Carniani}, Stefano and {Tacchella}, Sandro and {Baker}, William M. and {Arribas}, Santiago and {Nakajima}, Kimihiko and {Eisenstein}, Daniel J. and {Bunker}, Andrew J. and {Charlot}, St{\'e}phane and {Cresci}, Giovanni and {Curti}, Mirko and {Curtis-Lake}, Emma and {de Graaff}, Anna and {Egami}, Eiichi and {Ji}, Zhiyuan and {Johnson}, Benjamin D. and {Kumari}, Nimisha and {Looser}, Tobias J. and {Maseda}, Michael and {Nelson}, Erica and {Robertson}, Brant and {Rodr{\'\i}guez Del Pino}, Bruno and {Sandles}, Lester and {Simmonds}, Charlotte and {Smit}, Renske and {Sun}, Fengwu and {Venturi}, Giacomo and {Williams}, Christina C. and {Willmer}, Christopher N.~A.},
        title = "{JADES. Possible Population III signatures at z = 10.6 in the halo of GN-z11}",
      journal = {\aap},
         year = 2024,
        month = jul,
       volume = {687},
          eid = {A67},
        pages = {A67},
          doi = {10.1051/0004-6361/202347087},
archivePrefix = {arXiv},
       eprint = {2306.00953},
 primaryClass = {astro-ph.GA},
       adsurl = {https://ui.adsabs.harvard.edu/abs/2024A&A...687A..67M}
}

@ARTICLE{Scholtz2024GNz11,
       author = {{Scholtz}, Jan and {Witten}, Callum and {Laporte}, Nicolas and {{\"U}bler}, Hannah and {Perna}, Michele and {Maiolino}, Roberto and {Arribas}, Santiago and {Baker}, William M. and {Bennett}, Jake S. and {D'Eugenio}, Francesco and {Simmonds}, Charlotte and {Tacchella}, Sandro and {Witstok}, Joris and {Bunker}, Andrew J. and {Carniani}, Stefano and {Charlot}, St{\'e}phane and {Cresci}, Giovanni and {Curtis-Lake}, Emma and {Eisenstein}, Daniel J. and {Kumari}, Nimisha and {Robertson}, Brant and {Rodr{\'\i}guez Del Pino}, Bruno and {Smit}, Renske and {Venturi}, Giacomo and {Williams}, Christina C. and {Willmer}, Christopher N.~A.},
        title = "{GN-z11: The environment of an active galactic nucleus at z = 10.603. New insights into the most distant Ly{\ensuremath{\alpha}} detection}",
      journal = {\aap},
         year = 2024,
        month = jul,
       volume = {687},
          eid = {A283},
        pages = {A283},
          doi = {10.1051/0004-6361/202347187},
archivePrefix = {arXiv},
       eprint = {2306.09142},
 primaryClass = {astro-ph.GA},
       adsurl = {https://ui.adsabs.harvard.edu/abs/2024A&A...687A.283S}
}

@ARTICLE{Tacchella2023,
       author = {{Tacchella}, Sandro and {Eisenstein}, Daniel J. and {Hainline}, Kevin and {Johnson}, Benjamin D. and {Baker}, William M. and {Helton}, Jakob M. and {Robertson}, Brant and {Suess}, Katherine A. and {Chen}, Zuyi and {Nelson}, Erica and {Pusk{\'a}s}, D{\'a}vid and {Sun}, Fengwu and {Alberts}, Stacey and {Egami}, Eiichi and {Hausen}, Ryan and {Rieke}, George and {Rieke}, Marcia and {Shivaei}, Irene and {Williams}, Christina C. and {Willmer}, Christopher N.~A. and {Bunker}, Andrew and {Cameron}, Alex J. and {Carniani}, Stefano and {Charlot}, Stephane and {Curti}, Mirko and {Curtis-Lake}, Emma and {Looser}, Tobias J. and {Maiolino}, Roberto and {Maseda}, Michael V. and {Rawle}, Tim and {Rix}, Hans-Walter and {Smit}, Renske and {{\"U}bler}, Hannah and {Willott}, Chris and {Witstok}, Joris and {Baum}, Stefi and {Bhatawdekar}, Rachana and {Boyett}, Kristan and {Danhaive}, A. Lola and {de Graaff}, Anna and {Endsley}, Ryan and {Ji}, Zhiyuan and {Lyu}, Jianwei and {Sandles}, Lester and {Saxena}, Aayush and {Scholtz}, Jan and {Topping}, Michael W. and {Whitler}, Lily},
        title = "{JADES Imaging of GN-z11: Revealing the Morphology and Environment of a Luminous Galaxy 430 Myr after the Big Bang}",
      journal = {\apj},
         year = 2023,
        month = jul,
       volume = {952},
       number = {1},
          eid = {74},
        pages = {74},
          doi = {10.3847/1538-4357/acdbc6},
archivePrefix = {arXiv},
       eprint = {2302.07234},
 primaryClass = {astro-ph.GA},
       adsurl = {https://ui.adsabs.harvard.edu/abs/2023ApJ...952...74T}
}

@ARTICLE{Oesch2016,
       author = {{Oesch}, P.~A. and {Brammer}, G. and {van Dokkum}, P.~G. and {Illingworth}, G.~D. and {Bouwens}, R.~J. and {Labb{\'e}}, I. and {Franx}, M. and {Momcheva}, I. and {Ashby}, M.~L.~N. and {Fazio}, G.~G. and {Gonzalez}, V. and {Holden}, B. and {Magee}, D. and {Skelton}, R.~E. and {Smit}, R. and {Spitler}, L.~R. and {Trenti}, M. and {Willner}, S.~P.},
        title = "{A Remarkably Luminous Galaxy at z=11.1 Measured with Hubble Space Telescope Grism Spectroscopy}",
      journal = {\apj},
         year = 2016,
        month = mar,
       volume = {819},
       number = {2},
          eid = {129},
        pages = {129},
          doi = {10.3847/0004-637X/819/2/129},
archivePrefix = {arXiv},
       eprint = {1603.00461},
 primaryClass = {astro-ph.GA},
       adsurl = {https://ui.adsabs.harvard.edu/abs/2016ApJ...819..129O}
}

@ARTICLE{Maiolino2024JADES,
       author = {{Maiolino}, Roberto and {Scholtz}, Jan and {Curtis-Lake}, Emma and {Carniani}, Stefano and {Baker}, William and {de Graaff}, Anna and {Tacchella}, Sandro and {{\"U}bler}, Hannah and {D'Eugenio}, Francesco and {Witstok}, Joris and {Curti}, Mirko and {Arribas}, Santiago and {Bunker}, Andrew J. and {Charlot}, St{\'e}phane and {Chevallard}, Jacopo and {Eisenstein}, Daniel J. and {Egami}, Eiichi and {Ji}, Zhiyuan and {Jones}, Gareth C. and {Lyu}, Jianwei and {Rawle}, Tim and {Robertson}, Brant and {Rujopakarn}, Wiphu and {Perna}, Michele and {Sun}, Fengwu and {Venturi}, Giacomo and {Williams}, Christina C. and {Willott}, Chris},
        title = "{JADES: The diverse population of infant black holes at 4 < z < 11: Merging, tiny, poor, but mighty}",
      journal = {\aap},
         year = 2024,
        month = nov,
       volume = {691},
          eid = {A145},
        pages = {A145},
          doi = {10.1051/0004-6361/202347640},
archivePrefix = {arXiv},
       eprint = {2308.01230},
 primaryClass = {astro-ph.GA},
       adsurl = {https://ui.adsabs.harvard.edu/abs/2024A&A...691A.145M}
}

@ARTICLE{Juodzbalis2025,
       author = {{Juod{\v{z}}balis}, Ignas and {Maiolino}, Roberto and {Baker}, William M. and {Lake}, Emma Curtis and {Scholtz}, Jan and {D'Eugenio}, Francesco and {Trefoloni}, Bartolomeo and {Isobe}, Yuki and {Tacchella}, Sandro and {Bunker}, Andrew J. and {Carniani}, Stefano and {Charlot}, St{\'e}phane and {Jones}, Gareth C. and {Parlanti}, Eleonora and {Perna}, Michele and {Rinaldi}, Pierluigi and {Robertson}, Brant and {{\"U}bler}, Hannah and {Venturi}, Giacomo and {Willott}, Chris},
        title = "{JADES: comprehensive census of broad-line AGN from Reionization to Cosmic Noon revealed by JWST}",
      journal = {arXiv e-prints},
         year = 2025,
        month = apr,
          eid = {arXiv:2504.03551},
        pages = {arXiv:2504.03551},
          doi = {10.48550/arXiv.2504.03551},
archivePrefix = {arXiv},
       eprint = {2504.03551},
 primaryClass = {astro-ph.GA},
       adsurl = {https://ui.adsabs.harvard.edu/abs/2025arXiv250403551J}
}

@ARTICLE{Kocevski2025,
       author = {{Kocevski}, Dale D. and {Finkelstein}, Steven L. and {Barro}, Guillermo and {Taylor}, Anthony J. and {Calabr{\`o}}, Antonello and {Laloux}, Brivael and {Buchner}, Johannes and {Trump}, Jonathan R. and {Leung}, Gene C.~K. and {Yang}, Guang and {Dickinson}, Mark and {P{\'e}rez-Gonz{\'a}lez}, Pablo G. and {Pacucci}, Fabio and {Inayoshi}, Kohei and {Somerville}, Rachel S. and {McGrath}, Elizabeth J. and {Akins}, Hollis B. and {Bagley}, Micaela B. and {Bowler}, Rebecca A.~A. and {Bisigello}, Laura and {Carnall}, Adam and {Casey}, Caitlin M. and {Cheng}, Yingjie and {Cleri}, Nikko J. and {Costantin}, Luca and {Cullen}, Fergus and {Davis}, Kelcey and {Donnan}, Callum T. and {Dunlop}, James S. and {Ellis}, Richard S. and {Ferguson}, Henry C. and {Fujimoto}, Seiji and {Fontana}, Adriano and {Giavalisco}, Mauro and {Grazian}, Andrea and {Grogin}, Norman A. and {Hathi}, Nimish P. and {Hirschmann}, Michaela and {Huertas-Company}, Marc and {Holwerda}, Benne W. and {Illingworth}, Garth and {Juneau}, St{\'e}phanie and {Kartaltepe}, Jeyhan S. and {Koekemoer}, Anton M. and {Li}, Wenxiu and {Lucas}, Ray A. and {Magee}, Dan and {Mason}, Charlotte and {McLeod}, Derek J. and {McLure}, Ross J. and {Napolitano}, Lorenzo and {Papovich}, Casey and {Pirzkal}, Nor and {Rodighiero}, Giulia and {Santini}, Paola and {Wilkins}, Stephen M. and {Yung}, L.~Y. Aaron},
        title = "{The Rise of Faint, Red Active Galactic Nuclei at z > 4: A Sample of Little Red Dots in the JWST Extragalactic Legacy Fields}",
      journal = {\apj},
         year = 2025,
        month = jun,
       volume = {986},
       number = {2},
          eid = {126},
        pages = {126},
          doi = {10.3847/1538-4357/adbc7d},
archivePrefix = {arXiv},
       eprint = {2404.03576},
 primaryClass = {astro-ph.GA},
       adsurl = {https://ui.adsabs.harvard.edu/abs/2025ApJ...986..126K}
}

@ARTICLE{Taylor2025,
       author = {{Taylor}, Anthony J. and {Finkelstein}, Steven L. and {Kocevski}, Dale D. and {Jeon}, Junehyoung and {Bromm}, Volker and {Amor{\'\i}n}, Ricardo O. and {Arrabal Haro}, Pablo and {Backhaus}, Bren E. and {Bagley}, Micaela B. and {Banados}, Eduardo and {Bhatawdekar}, Rachana and {Brooks}, Madisyn and {Calabr{\`o}}, Antonello and {Ch{\'a}vez Ortiz}, {\'O}scar A. and {Cheng}, Yingjie and {Cleri}, Nikko J. and {Cole}, Justin W. and {Davis}, Kelcey and {Dickinson}, Mark and {Donnan}, Callum and {Dunlop}, James S. and {Ellis}, Richard S. and {Fern{\'a}ndez}, Vital and {Fontana}, Adriano and {Fujimoto}, Seiji and {Giavalisco}, Mauro and {Grazian}, Andrea and {Guo}, Jingsong and {Hathi}, Nimish P. and {Holwerda}, Benne W. and {Hirschmann}, Michaela and {Inayoshi}, Kohei and {Kartaltepe}, Jeyhan S. and {Khusanova}, Yana and {Koekemoer}, Anton M. and {Kokorev}, Vasily and {Larson}, Rebecca L. and {Leung}, Gene C.~K. and {Lucas}, Ray A. and {McLeod}, Derek J. and {Napolitano}, Lorenzo and {Onoue}, Masafusa and {Pacucci}, Fabio and {Papovich}, Casey and {P{\'e}rez-Gonz{\'a}lez}, Pablo G. and {Pirzkal}, Nor and {Somerville}, Rachel S. and {Trump}, Jonathan R. and {Wilkins}, Stephen M. and {Yung}, L.~Y. Aaron and {Zhang}, Haowen},
        title = "{Broad-line AGNs at 3.5 < z < 6: The Black Hole Mass Function and a Connection with Little Red Dots}",
      journal = {\apj},
         year = 2025,
        month = jun,
       volume = {986},
       number = {2},
          eid = {165},
        pages = {165},
          doi = {10.3847/1538-4357/add15b},
archivePrefix = {arXiv},
       eprint = {2409.06772},
 primaryClass = {astro-ph.GA},
       adsurl = {https://ui.adsabs.harvard.edu/abs/2025ApJ...986..165T}
}

@ARTICLE{Naidu25,
       author = {{Naidu}, Rohan P. and {Oesch}, Pascal A. and {Brammer}, Gabriel and {Weibel}, Andrea and {Li}, Yijia and {Matthee}, Jorryt and {Chisholm}, John and {Pollock}, Clara L. and {Heintz}, Kasper E. and {Johnson}, Benjamin D. and {Shen}, Xuejian and {Hviding}, Raphael E. and {Leja}, Joel and {Tacchella}, Sandro and {Ganguly}, Arpita and {Witten}, Callum and {Atek}, Hakim and {Belli}, Sirio and {Bose}, Sownak and {Bouwens}, Rychard and {Dayal}, Pratika and {Decarli}, Roberto and {de Graaff}, Anna and {Fudamoto}, Yoshinobu and {Giovinazzo}, Emma and {Greene}, Jenny E. and {Illingworth}, Garth and {Inoue}, Akio K. and {Kane}, Sarah G. and {Labbe}, Ivo and {Leonova}, Ecaterina and {Marques-Chaves}, Rui and {Meyer}, Romain A. and {Nelson}, Erica J. and {Roberts-Borsani}, Guido and {Schaerer}, Daniel and {Simcoe}, Robert A. and {Stefanon}, Mauro and {Sugahara}, Yuma and {Toft}, Sune and {van der Wel}, Arjen and {van Dokkum}, Pieter and {Walter}, Fabian and {Watson}, Darach and {Weaver}, John R. and {Whitaker}, Katherine E.},
        title = "{A Cosmic Miracle: A Remarkably Luminous Galaxy at $z_{\rm{spec}}=14.44$ Confirmed with JWST}",
      journal = {arXiv e-prints},
         year = 2025,
        month = may,
          eid = {arXiv:2505.11263},
        pages = {arXiv:2505.11263},
          doi = {10.48550/arXiv.2505.11263},
archivePrefix = {arXiv},
       eprint = {2505.11263},
 primaryClass = {astro-ph.GA},
       adsurl = {https://ui.adsabs.harvard.edu/abs/2025arXiv250511263N}
}

@ARTICLE{SS73,
       author = {{Shakura}, N.~I. and {Sunyaev}, R.~A.},
        title = "{Black holes in binary systems. Observational appearance.}",
      journal = {\aap},
         year = 1973,
        month = jan,
       volume = {24},
        pages = {337-355},
       adsurl = {https://ui.adsabs.harvard.edu/abs/1973A&A....24..337S}
}

@INPROCEEDINGS{NT73,
       author = {{Novikov}, I.~D. and {Thorne}, K.~S.},
        title = "{Astrophysics of black holes.}",
    booktitle = {Black Holes (Les Astres Occlus)},
         year = 1973,
       editor = {{Dewitt}, C. and {Dewitt}, B.~S.},
        month = jan,
        pages = {343-450},
       adsurl = {https://ui.adsabs.harvard.edu/abs/1973blho.conf..343N}
}

@ARTICLE{Li05,
       author = {{Li}, Li-Xin and {Zimmerman}, Erik R. and {Narayan}, Ramesh and {McClintock}, Jeffrey E.},
        title = "{Multitemperature Blackbody Spectrum of a Thin Accretion Disk around a Kerr Black Hole: Model Computations and Comparison with Observations}",
      journal = {\apjs},
         year = 2005,
        month = apr,
       volume = {157},
       number = {2},
        pages = {335-370},
          doi = {10.1086/428089},
archivePrefix = {arXiv},
       eprint = {astro-ph/0411583},
 primaryClass = {astro-ph},
       adsurl = {https://ui.adsabs.harvard.edu/abs/2005ApJS..157..335L}
}

@ARTICLE{Maiolino25,
       author = {{Maiolino}, Roberto and {Uebler}, Hannah and {D'Eugenio}, Francesco and {Scholtz}, Jan and {Juodzbalis}, Ignas and {Ji}, Xihan and {Perna}, Michele and {Bromm}, Volker and {Dayal}, Pratika and {Koudmani}, Sophie and {Liu}, Boyuan and {Schneider}, Raffaella and {Sijacki}, Debora and {Valiante}, Rosa and {Trinca}, Alessandro and {Zhang}, Saiyang and {Volonteri}, Marta and {Inayoshi}, Kohei and {Carniani}, Stefano and {Nakajima}, Kimihiko and {Isobe}, Yuki and {Witstok}, Joris and {Jones}, Gareth C. and {Tacchella}, Sandro and {Arribas}, Santiago and {Bunker}, Andrew and {Cataldi}, Elisa and {Charlot}, Stephane and {Cresci}, Giovanni and {Curti}, Mirko and {Fabian}, Andrew C. and {Katz}, Harley and {Kumari}, Nimisha and {Laporte}, Nicolas and {Mazzolari}, Giovanni and {Robertson}, Brant and {Sun}, Fengwu and {Rodriguez Del Pino}, Bruno and {Venturi}, Giacomo},
        title = "{A black hole in a near-pristine galaxy 700 million years after the Big Bang}",
      journal = {arXiv e-prints},
         year = 2025,
        month = may,
          eid = {arXiv:2505.22567},
        pages = {arXiv:2505.22567},
          doi = {10.48550/arXiv.2505.22567},
archivePrefix = {arXiv},
       eprint = {2505.22567},
 primaryClass = {astro-ph.GA},
       adsurl = {https://ui.adsabs.harvard.edu/abs/2025arXiv250522567M}
}

@PROCEEDING{arnaud96,
   author = {{Arnaud}, K.~A.},
    title = "{XSPEC: The First Ten Years}",
booktitle = {Astronomical Data Analysis Software and Systems V},
     year = 1996,
   series = {Astronomical Society of the Pacific Conference Series},
   volume = 101,
   editor = {{Jacoby}, G.~H. and {Barnes}, J.},
    pages = {17},
   adsurl = {http://adsabs.harvard.edu/abs/1996ASPC..101...17A}
}

@ARTICLE{Coughlin24,
       author = {{Coughlin}, Eric R. and {Begelman}, Mitchell C.},
        title = "{Quasi-stars as a Means of Rapid Black Hole Growth in the Early Universe}",
      journal = {\apj},
         year = 2024,
        month = aug,
       volume = {970},
       number = {2},
          eid = {158},
        pages = {158},
          doi = {10.3847/1538-4357/ad5723},
archivePrefix = {arXiv},
       eprint = {2405.00084},
 primaryClass = {astro-ph.GA},
       adsurl = {https://ui.adsabs.harvard.edu/abs/2024ApJ...970..158C}
}

@article{castellano24,
doi = {10.3847/1538-4357/ad5f88},
url = {https://dx.doi.org/10.3847/1538-4357/ad5f88},
year = {2024},
month = {sep},
publisher = {The American Astronomical Society},
volume = {972},
number = {2},
pages = {143},
author = {Castellano, Marco and Napolitano, Lorenzo and Fontana, Adriano and Roberts-Borsani, Guido and Treu, Tommaso and Vanzella, Eros and Zavala, Jorge A. and Arrabal Haro, Pablo and Calabrò, Antonello and Llerena, Mario and Mascia, Sara and Merlin, Emiliano and Paris, Diego and Pentericci, Laura and Santini, Paola and Bakx, Tom J. L. C. and Bergamini, Pietro and Cupani, Guido and Dickinson, Mark and Filippenko, Alexei V. and Glazebrook, Karl and Grillo, Claudio and Kelly, Patrick L. and Malkan, Matthew A. and Mason, Charlotte A. and Morishita, Takahiro and Nanayakkara, Themiya and Rosati, Piero and Sani, Eleonora and Wang, Xin and Yoon, Ilsang},
title = {JWST NIRSpec Spectroscopy of the Remarkable Bright Galaxy GHZ2/GLASS-z12 at Redshift 12.34},
journal = {The Astrophysical Journal}
}

@ARTICLE{Netzer19,
       author = {{Netzer}, Hagai},
        title = "{Bolometric correction factors for active galactic nuclei}",
      journal = {\mnras},
         year = 2019,
        month = oct,
       volume = {488},
       number = {4},
        pages = {5185-5191},
          doi = {10.1093/mnras/stz2016},
archivePrefix = {arXiv},
       eprint = {1907.09534},
 primaryClass = {astro-ph.GA},
       adsurl = {https://ui.adsabs.harvard.edu/abs/2019MNRAS.488.5185N}
}
 




\newpage

\appendix 
\section{The UV Spectrum of  GN=z11 compared with that of  the   low redshift AGN 1H 0707-495} 

In Fig. 4 we compare both the rest-frame UV spectrum of GN-z11 from JWST and the HST STIS spectrum of the Narrow-Line Seyfert 1 galaxy, 1H0707-495 at $z=0.04$ \citep{Leighly04} with the composite intrinsic quasar spectrum of \citep{Temple21}. The solid line shown in the lower panel is the best-fit intrinsic disc model for 1H0707 after removing the Galactic column density of $N_{\rm H}=2.6\times10^{20}$\,cm$^{-2}$ and extinction of $E(B-V)=0.113$, which has the effect of flattening the spectrum.

\begin{figure}
    \centering
    \includegraphics[width=1.0\linewidth]{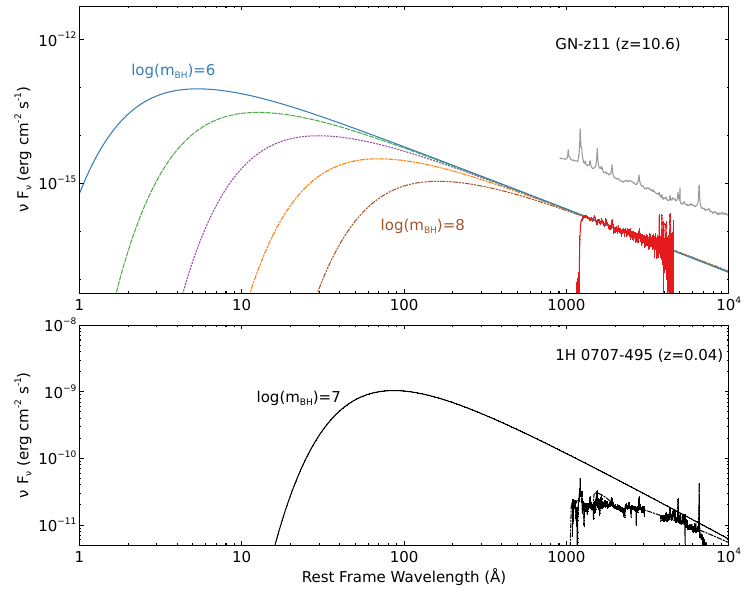}
    \caption{Top: the same best-fit disc thermal emission model as in the bottom right panel of Fig.\,\ref{pic_jwst_fit3}. The solid line shows the best-fit model for a range of BH masses from $10^{6}$ to $10^{8}$ solar masses. The grey line is the average $z=0$ intrinsic quasar spectrum calculated using QSOGEN \citep{Temple21}. Note that it compares well to the accretion disc. Bottom: the HST (UV) and UKST (optical) spectra of a local NLS1 AGN 1H 0707$-$495 ($z=0.04$). Its UV and optical continuum is consistent with an absorbed disc thermal emission ($a_{*}=0.99$, $i=60^{\circ}$, $m_{\rm BH}=10^{7}$, dash-dotted line), after removing the effects of dust extinction. }
    \label{pic_compare3}
\end{figure}



\bsp	
\label{lastpage}
\end{document}